# The Dirac equation in Rindler space: A pedagogical introduction

David McMahon
Sandia National Laboratories
Albuquerque, NM 87185
Email: dmmcmah@sandia.gov

Paul M. Alsing
Department of Physics and Astronomy, University of New Mexico,
Albuquerque, NM 87131
alsing@hpc.unm.edu

Pedro Embid
Department of Mathematics and Statistics, University of New Mexico,
Albuquerque, NM 87131

December 28, 2005

**Abstract**

A pedagogical introduction to the Dirac equation for massive particles in Rindler space is presented. The spin connection coefficients are explicitly derived using techniques from general relativity. We then apply the Lagrange-Green identity to greatly simplify calculation of the inner products needed to normalize the states. Finally, the Bogolubov coefficients relating the Rindler and Minkowski modes are derived in an intuitive manner. These derivations are useful for students interested in learning about quantum field theory in a curved space-time.

## 1 Introduction

One of the most dramatic results in the early attempts to marry quantum field theory to general relativity was the discovery by Hawking that black holes radiate [1]. This interesting result was soon followed by the discovery by Unruh [2] that a uniformly accelerated detector would be immersed in a bath of thermal radiation with temperature $T = \hbar a / 2\pi$, where $a$ is the acceleration of the observer ($k = c = 1$). Moreover, an accelerated observer in flat space-time has his or her own event horizon. The result of Unruh can be encapsulated by saying each observer has his or her own vacuum.

These celebrated results, together with a bit of simplicity that flat space-time brings, have made Rindler space a very interesting arena within which to study quantum field theory. While most treatments have focused on the scalar field, one effort many years ago briefly showed that a uniformly accelerated observer would see a flux of massive Dirac particles [3].

Due to the recent interest in quantum information theory in a relativistic context, in particular the consideration of teleportation with an accelerated observer [4], in this primarily pedagogical paper we recall the main results associated with Dirac particles in Rindler space. Our motivation is two fold. First, this paper is designed to assist students interested in studying quantum field theory in a curved space-time. In addition, we aim to put the necessary machinery on the table so that a later work can consider quantum information theory with accelerated observers using a complete relativistic framework.

The outline of this paper is as follows. After briefly describing Rindler coordinates and the form of the Dirac equation in a curved space-time, we derive the *spin connection coefficients* by considering the geodesic equation. In the next section, we address the crucial problem of normalizing the states, which are given in terms of Hankel functions. This is a non-trivial problem. This paper addresses that fact by applying the *Lagrange-Green identity* to greatly simplify the calculation. To our knowledge, this approach has not been used in this case. Finally, after normalizing the states, we show how to write the Minkowski wave functions in terms of the Rindler wave functions and vice versa. This allows us to conclude the paper by deriving the Bogolubov coefficients.

## 2 Rindler Coordinates

In this paper we consider the familiar case of a uniformly accelerated observer in Minkowski space-time. As shown in Fig. 1, we consider two-dimensional Minkowski space with coordinates $(t,z)$ divided into four regions. These are the right Rindler wedge (region I), the left Rindler wedge (region II) and the future (F) and past (P). Setting $\hbar = c = 1$, we define Rindler coordinates $(v,u)$ which can be related to Minkowski coordinates in the following way

$$\left.\begin{array}{ll} t = u\sinh v & z = u\cosh v \\ v = \tanh^{-1}(t/z) & u = \text{sgn}(z)\sqrt{z^2 - t^2} \end{array}\right\} \text{ in regions I, II}$$
$$\left.\begin{array}{ll} t = u\cosh v & z = u\sinh v \\ v = \tanh^{-1}(z/t) & u = \text{sgn}(t)\sqrt{t^2 - z^2} \end{array}\right\} \text{ in regions F, P} \tag{1}$$

Since the calculations described in this paper will be similar for all four regions, we focus on region I, which we denote as *Rindler space*. In this region, starting with the two-dimensional Minkowski line element $ds^2 = dt^2 - dz^2$ and using (1), it is easy to show that the line element takes the form

$$ds^2 = u^2 dv^2 - du^2 \tag{2}$$

Note that $v$ is a timelike coordinate and $u$ is a spacelike coordinate-in region I. This will change in the other regions. Using this line element we can read off the components of the metric tensor. These can be arranged in a matrix as follows

$$g_{mn} = \begin{pmatrix} u & 0 \\ 0 & -1 \end{pmatrix} \tag{3}$$

The components of $g^{mn}$ can be found easily by inverting (3). We find that

$$g^{mn} = \begin{pmatrix} 1/u & 0 \\ 0 & -1 \end{pmatrix} \qquad (4)$$

In the next section, we proceed to derive the form of the Dirac equation in Rindler space.

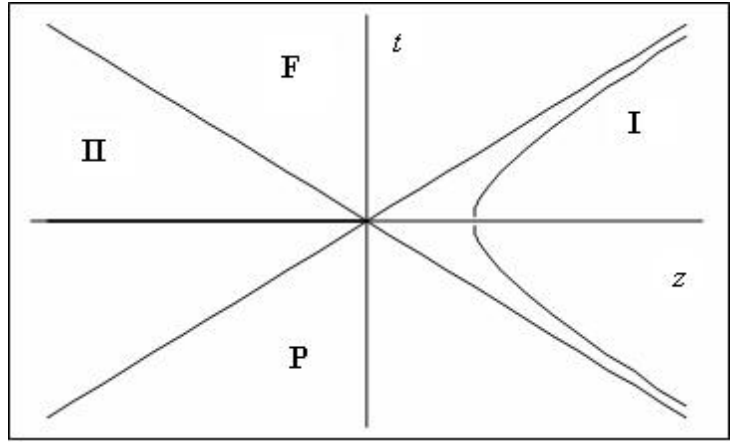

Fig. 1. The division of Minkowski space into four regions, the right Rindler wedge (I), the left Rindler wedge (II), future (F), and past (P). The hyperbola is the worldline of an accelerated observer.

## 3 The Dirac equation in Rindler space

Given general curvilinear coordinates with metric $g_{mn}$, the Dirac equation is written as[5]

$$\left[ ig^m \left( \frac{\partial}{\partial x^m} + \Gamma_m \right) - m \right] \psi = 0 \qquad (5)$$

where the $\Gamma_m$ are known as the *spin connection coefficients*. These are $4 \times 4$ matrices that can be written in terms of the Gamma matrices in the following way

$$\Gamma_m = \frac{1}{4} g_n \left( \frac{\partial g^n}{\partial x^m} + \Gamma^n{}_{lm} g^l \right) \qquad (6)$$

Many readers will recognize the $\Gamma^n{}_{lm}$ as the famed *Christoffel symbols* familiar from general relativity. Note that the $g^m$ shown here are not the "usual" Dirac matrices from flat space, instead they depend on the metric in the following way

$$g^m g^n + g^n g^m = 2g^{mn} \tag{7}$$

We can use this equation to relate these gamma matrices to the usual flat space variety. If we denote the flat space gamma matrices by $\bar{g}^m$, using the fact that $(\bar{g}^0)^2 = +1$ and $(\bar{g}^3)^2 = -1$ together with (7) and (3), we immediately deduce that

$$g_0^2 = g_{00}^2 = u^2 (\bar{g}_0)^2, \Rightarrow g_0 = u\bar{g}_0 \tag{8}$$

$$g_3^2 = g_{33}^2 = -1 = (\bar{g}_3)^2, \Rightarrow g_3 = \bar{g}_3 \tag{9}$$

Using $g^{mn}$ to raise indices it can be shown that

$$g^0 = \frac{1}{u}\bar{g}^0, \quad g^3 = \bar{g}^3 \tag{10}$$

Our next task on the road to computing the spin connection coefficients is to turn to the problem of finding the form of the Christoffel symbols. To do this we begin with a brief study of the geodesic equation. General relativity tells us that the paths of particles moving without the influence of any external forces are determined by the following equation

$$\ddot{x}^m + \Gamma^n{}_{lm} \dot{x}^l \dot{x}^n = 0 \tag{11}$$

where the derivatives indicated by a dot are taken with respect to an affine parameter. The explicit form of the Christoffel symbols and therefore of the spin connection coefficients can be determined in a fairly painless fashion by applying a Lagrangian procedure [6]. We start by defining a quantity which we label as $K$

$$K = \frac{1}{2} g_{mn} \dot{x}^m \dot{x}^n \tag{12}$$

In the case of Rindler coordinates in region I, we have [7]

$$K = \frac{1}{2}\left[ u^2 \dot{v}^2 - \dot{u}^2 \right] \tag{13}$$

Calling the affine parameter $t$, we can use $K$ to rewrite the geodesic equation, and then compare it with (11) to simply read off the Christoffel symbols. We have

$$\frac{\partial K}{\partial x^m} - \frac{d}{dt}\left( \frac{\partial K}{\partial \dot{x}^m} \right) = 0 \tag{14}$$

First we consider the $v$ coordinate. Setting $x^m = v$ we find

$$\frac{\partial K}{\partial v} = 0, \quad \Rightarrow -\frac{d}{dt}\left(\frac{\partial K}{\partial \dot{v}}\right) = 0 \qquad (15)$$

Using the definition of $K$ given in (13) this leads to the following equation

$$\frac{d}{dt}\left(\frac{\partial K}{\partial \dot{v}}\right) = 2u\dot{u}\dot{v} + u^2\ddot{v} \qquad (16)$$

Rearranging terms and setting equal to zero by virtue of (15) gives

$$\ddot{v} + \frac{2}{u}\dot{u}\dot{v} = 0 \qquad (17)$$

Comparison with the geodesic equation (11) allows us to read off the values of the following Christoffel symbols

$$\Gamma^v_{uv} = \Gamma^v_{vu} = \frac{2}{u} \qquad (18)$$

A similar procedure using $x^m = u$ shows that the only other non-zero Christoffel symbol is given by

$$\Gamma^u_{vv} = u \qquad (19)$$

Now that we have explicit expressions for the Christoffel symbols, we can write down the spin connection coefficients using (6). We will calculate one example explicitly. The equation for $\Gamma_v$ is given by

$$\Gamma_v = \frac{1}{4}g_0\frac{\partial g^0}{\partial v} + \frac{1}{4}g_0\Gamma^v_{1v}g^1 \qquad (20)$$

We are following the summation convention, so the index $1$ is to be summed over all coordinates. Writing this result in terms of the flat space gamma matrices and explicitly writing out the sum, we have

$$\begin{aligned}\Gamma_v &= \frac{1}{4}u\bar{g}_0\Gamma^v_{uv}\bar{g}^3 + \frac{1}{4}\bar{g}_3\Gamma^u_{vv}\frac{1}{u}\bar{g}^0 \\ &= \frac{1}{4}\left(\bar{g}^0\bar{g}^3 + \bar{g}^3\bar{g}^0\right) \\ &= -\frac{1}{2}\bar{g}^0\bar{g}^3\end{aligned} \qquad (21)$$

A similar procedure can be used to show that $\Gamma_u = 0$. We now have everything we need to write down the full Dirac equation. Using (10) together with (21) in (5) we find that

$$0 = i\bar{g}^m\left(\frac{\partial}{\partial x^m} + \Gamma_m\right)y - my$$

$$= \frac{i}{u}\bar{g}^0\frac{\partial y}{\partial v} + i\bar{g}^3\frac{\partial y}{\partial u} - \frac{i}{2u}\bar{g}^0\left(\bar{g}^0\bar{g}^3\right)y - my \tag{22}$$

Using $\left(\bar{g}^0\right)^2 = +1$ and rearranging terms gives

$$i\frac{\partial y}{\partial v} = -iu\bar{g}^0\bar{g}^3\frac{\partial y}{\partial u} + \frac{i}{2}\bar{g}^0\bar{g}^3 y + m\bar{g}^0 y \tag{23}$$

Now we define $\bar{g}^0\bar{g}^3 = a_3$ and denote $\bar{g}^0$ by $b$ to write the Dirac equation as

$$i\frac{\partial y}{\partial v} = u\left[-ia_3\left(\frac{\partial}{\partial u} + \frac{1}{2u}\right) + bm\right]y \tag{24}$$

In this paper we use the representation where

$$a_3 = \begin{pmatrix} 0 & 0 & 1 & 0 \\ 0 & 0 & 0 & -1 \\ 1 & 0 & 0 & 0 \\ 0 & -1 & 0 & 0 \end{pmatrix}, \quad b = \begin{pmatrix} 1 & 0 & 0 & 0 \\ 0 & 1 & 0 & 0 \\ 0 & 0 & -1 & 0 \\ 0 & 0 & 0 & -1 \end{pmatrix} \tag{25}$$

Our next task is to write down the solutions to (24) and normalize the states.

## 3 Normalization with the Lagrange-Green Identity

The ultimate goal of the present exercise is to relate the complete Minkowski wave function to the Rindler wave functions in all four regions. In order to do this properly we will need to normalize the states. This turns out to be fairly mathematically difficult and may even appear a bit mysterious to most readers. However, as we'll see below, this process is greatly simplified by applying the Lagrange-Green identity.

The first step is to look at (24) by thinking of the Dirac equation as a differential operator. We will call that operator $L$ and note by looking at the right hand side of the Dirac equation (24), we can write $L$ as

$$L = u\left[-ia_3\left(\frac{\partial}{\partial u} + \frac{1}{2u}\right) + bm\right] \tag{26}$$

Of course we can view the Dirac equation as an operator and look at the left side of (24) as well. In that case we can write

$$L = i\frac{\partial}{\partial v} \tag{27}$$

Stationary states will be of the form $\mathbf{y} = e^{-iwv}\mathbf{y}(u)$, in which case the application of (27) immediately yields

$$L\mathbf{y} = -iw\mathbf{y} \tag{28}$$

To proceed further we will need the explicit form of the states. In this paper we consider the spin-up case where $\mathbf{y}$ is a four component column vector given by[3]

$$\mathbf{y}_w(u,v) = e^{-iwv}\begin{pmatrix} f_w^-(mu) - f_w^+(mu) \\ 0 \\ f_w^-(mu) + f_w^+(mu) \\ 0 \end{pmatrix} \tag{29}$$

The $f_w^\pm(mu)$ satisfy the following differential equation

$$\left(u\frac{d}{du}u\frac{d}{du}\right)f_w^\pm(mu) = \left[m^2 u^2 - \left(w \mp \frac{i}{2}\right)^2\right]f_w^\pm(mu) \tag{30}$$

Solutions of (30) are given in terms of *Hankel functions of the first kind*

$$f_w^\pm(mu) = H_{iw\pm 1/2}^{(1)}(imu) \tag{31}$$

In region I, a Dirac state $\mathbf{y} \in L^2([0,\infty), \mathbb{C}^4)$. Therefore the inner product between two states $\mathbf{y}$ and $\mathbf{f}$ is given by

$$\langle \mathbf{y} | \mathbf{f} \rangle = \int_0^\infty \mathbf{y}^\dagger \mathbf{f}\, du \tag{32}$$

In order to normalize the states $\mathbf{y}_w$ as given in (29) we will need to compute the inner product

$$\langle \mathbf{y}_w | \mathbf{y}_{w'} \rangle = \int_0^\infty \mathbf{y}_w^\dagger \mathbf{y}_{w'}\, du \tag{33}$$

We can calculate this integral by applying the Lagrange-Green identity (described in Appendix B). We begin by examining the *adjoint* of the differential operator given in (26). In particular, we wish to show that

$$(L\mathbf{y})^\dagger \mathbf{f} = \frac{\partial}{\partial u}\left(iu\mathbf{y}^\dagger \mathbf{a}_3 \mathbf{f}\right) + \mathbf{y}^\dagger (L\mathbf{f}) \tag{34}$$

This relation corresponds to that described by (90) in the appendix. In due course we will see the value of this relation by noting that when considering inner products of states, the presence of the term $\frac{\partial}{\partial u}\left(iu\mathbf{y}^\dagger \mathbf{a}_3 \mathbf{f}\right)$ will allow us to write down the results of the relevant integrals almost by inspection.

Let's consider the expression $(L\mathbf{y})^\dagger \mathbf{f}$ in detail. Using (24) and (26), we see that this is

$$\left(-i\mathbf{a}_3 u \frac{\partial \mathbf{y}}{\partial u} + \frac{\mathbf{a}_3}{2u}\mathbf{y} + bm\mathbf{y}\right)^\dagger \mathbf{f} = \left(-i\mathbf{a}_3 u \frac{\partial \mathbf{y}}{\partial u}\right)^\dagger \mathbf{f} + \left(\frac{\mathbf{a}_3}{2u}\mathbf{y}\right)^\dagger \mathbf{f} + (bm\mathbf{y})^\dagger \mathbf{f} \tag{35}$$

We approach the goal of rewriting this expression so that $L$ is applied to $\mathbf{f}$ by considering each piece on the right hand side of (35) in turn. Beginning with the first term, we have

$$\left(-i\mathbf{a}_3 u \frac{\partial \mathbf{y}}{\partial u}\right)^\dagger \mathbf{f} = \left(iu \frac{\partial \mathbf{y}^\dagger}{\partial u} \mathbf{a}_3^\dagger\right) \mathbf{f} = \frac{\partial \mathbf{y}^\dagger}{\partial u}(i\mathbf{a}_3 u \mathbf{f}) \tag{36}$$

However, notice that

$$\frac{\partial}{\partial u}\left(\mathbf{y}^\dagger i\mathbf{a}_3 u \mathbf{f}\right) = \frac{\partial \mathbf{y}^\dagger}{\partial u}(i\mathbf{a}_3 u \mathbf{f}) + \mathbf{y}^\dagger i\mathbf{a}_3 \mathbf{f} + \mathbf{y}^\dagger i\mathbf{a}_3 u \frac{\partial \mathbf{f}}{\partial u} = \frac{\partial \mathbf{y}^\dagger}{\partial u}(i\mathbf{a}_3 u \mathbf{f}) + \mathbf{y}^\dagger \left(i\mathbf{a}_3 \frac{\partial(u\mathbf{f})}{\partial u}\right) \tag{37}$$

And so we can write

$$\left(-i\mathbf{a}_3 u \frac{\partial \mathbf{y}}{\partial u}\right)^\dagger \mathbf{f} = \frac{\partial}{\partial u}\left(\mathbf{y}^\dagger i\mathbf{a}_3 u \mathbf{f}\right) - \mathbf{y}^\dagger \left(i\mathbf{a}_3 \frac{\partial(u\mathbf{f})}{\partial u}\right) \tag{38}$$

Turning to the second term in (35) we have

$$\left(-i\frac{\mathbf{a}_3}{2}\mathbf{y}\right)^\dagger \mathbf{f} = \mathbf{y}^\dagger \left(i\frac{\mathbf{a}_3}{2}\mathbf{f}\right) \tag{39}$$

Finally, for the last term we find

$$(bm\mathbf{y})^\dagger \mathbf{f} = \mathbf{y}^\dagger m\mathbf{b}\mathbf{f} \tag{40}$$

Putting these results together gives

$$(L\mathbf{y})^\dagger \mathbf{f} = \frac{\partial}{\partial u}(i\mathbf{y}^\dagger \mathbf{a}_3 u \mathbf{f}) - \mathbf{y}^\dagger \left(i\mathbf{a}_3 u \frac{\partial \mathbf{f}}{\partial u}\right) - \mathbf{y}^\dagger (i\mathbf{a}_3 \mathbf{f})$$
$$+ \mathbf{y}^\dagger \left(i\frac{\mathbf{a}_3}{2}\mathbf{f}\right) + \mathbf{y}^\dagger m\mathbf{b}\mathbf{f} \tag{41}$$
$$= \frac{\partial}{\partial u}(i\mathbf{y}^\dagger \mathbf{a}_3 u \mathbf{f}) + \mathbf{y}^\dagger \left[-i\mathbf{a}_3 u \frac{\partial \mathbf{f}}{\partial u} - i\frac{\mathbf{a}_3}{2}\mathbf{f} + m\mathbf{b}\mathbf{f}\right]$$

However, looking back at the definition of $L$, we have

$$-i\mathbf{a}_3 u \frac{\partial \mathbf{f}}{\partial u} - i\frac{\mathbf{a}_3}{2}\mathbf{f} + m\mathbf{b}\mathbf{f} = L\mathbf{f} \tag{42}$$

Therefore we have shown that

$$(L\mathbf{y})^\dagger \mathbf{f} = \frac{\partial}{\partial u}(iu\mathbf{y}^\dagger \mathbf{a}_3 \mathbf{f}) + \mathbf{y}^\dagger (L\mathbf{f}) \tag{43}$$

Notice that using the result of the Lagrange-Green identity as applied to the Dirac equation, we have demonstrated that

$$\int_a^b (L\mathbf{y})^\dagger \mathbf{f} \, du = \mathbf{y}^\dagger (iu\mathbf{a}_3)\mathbf{f} \Big|_a^b + \int_a^b \mathbf{y}^\dagger (L\mathbf{f})^\dagger \, du \tag{44}$$

We'll see very shortly that it is now possible to compute the normalization integral given in (33) using (44).

Now we set $\mathbf{y} \to \mathbf{y}_w$ and $\mathbf{f} \to \mathbf{y}_{w'}$ and recall (28), where we found that $L\mathbf{y}_w = w\mathbf{y}_w$. Then the left hand side of (44) becomes

$$\int_0^\infty (L\mathbf{y}_w)^\dagger \mathbf{y}_{w'} \, du = \int_0^\infty (-iw\mathbf{y}_w)^\dagger \mathbf{y}_{w'} \, du = iw \int_0^\infty \mathbf{y}_w^\dagger \mathbf{y}_{w'} \, du \tag{45}$$

Now, making the substitutions $\mathbf{y} \to \mathbf{y}_w$ and $\mathbf{f} \to \mathbf{y}_{w'}$ in the last term of the right-hand side of (44) gives us $\int_a^b \mathbf{y}_w^\dagger (L\mathbf{y}_{w'})^\dagger \, du = iw' \int_0^\infty \mathbf{y}_w^\dagger \mathbf{y}_{w'} du$. Putting our results together (44) becomes

$$w\int_a^b \mathbf{y}_w^\dagger \mathbf{y}_{w'} du = \mathbf{y}_w^\dagger (iu\mathbf{a}_3)\mathbf{y}_{w'}\bigg|_a^b + w'\int_a^b \mathbf{y}_w^\dagger \mathbf{y}_{w'} du \tag{46}$$

Moving both integrals to the left hand side and dividing through by $w-w'$ gives the result we seek

$$\langle \mathbf{y}_w | \mathbf{y}_{w'}\rangle = \int_0^\infty \mathbf{y}_w^\dagger \mathbf{y}_{w'} du = \frac{iu}{w-w'} \mathbf{y}_w^\dagger \mathbf{a}_3 \mathbf{y}_{w'}\bigg|_0^\infty \tag{47}$$

To proceed we need to write down an explicit expression for $\frac{iu}{w-w'}\mathbf{y}_w^\dagger \mathbf{a}_3 \mathbf{y}_{w'}$. Recalling the form of the spin-up states given in (29), after some algebra we find that

$$\frac{i2u}{w-w'}\mathbf{y}_w^\dagger \mathbf{a}_3 \mathbf{y}_{w'} = \frac{i2u}{w-w'}\left[\left(f_w^-\right)^* f_{w'}^- - \left(f_w^+\right)^* f_{w'}^+\right]$$

$$= \frac{i2u}{w-w'}\left[H_{-iw-1/2}^{(1)}(-imu)H_{iw'-1/2}^{(1)}(imu) - H_{-iw+1/2}^{(1)}(-imu)H_{iw'+1/2}^{(1)}(imu)\right] \tag{48}$$

Now we proceed to evaluate this expression at the limits given in (47). This will be simplified somewhat due to the asymptotic form of the Hankel functions with large argument. For large $z$, $H_n^1(z)$ can be written as [8]

$$H_n^1(z) \to \sqrt{\frac{2}{pz}} \exp\left[i\left(z - \frac{np}{2} - \frac{p}{4}\right)\right] \tag{49}$$

The presence of the term $\sqrt{\frac{2}{pz}}$ ensures that as $z \to \infty$, these terms will vanish. Therefore (47) becomes

$$\int_0^\infty \mathbf{y}_w^\dagger \mathbf{y}_{w'} du = \frac{iu}{w-w'}\mathbf{y}_w^\dagger \mathbf{a}_3 \mathbf{y}_{w'}\bigg|_0^\infty = -\lim_{u\to 0}\frac{iu}{w-w'}\mathbf{y}_w^\dagger \mathbf{a}_3 \mathbf{y}_{w'} \tag{50}$$

Since both terms in (48) are similar, let's just focus on one of them. We consider

$$\lim_{u\to 0}\left(\frac{i2u}{w-w'}H_{-iw+1/2}^{(1)}(-imu)H_{iw'+1/2}^{(1)}(imu)\right) \tag{51}$$

Hankel functions can be written in terms of the modified Bessel functions of the second kind in the following way

$$K_n(z) = i\frac{p}{2}e^{inp/2}H_n^{(1)}(ze^{ip/2}) \tag{52}$$

This is convenient because for small $z$ the asymptotic form of the $K_n(z)$ is given by

$$K_n(z) \approx \frac{2^{n-1}\Gamma(n)}{z^n} \tag{53}$$

Inverting (52) and using (53) we can write (51) as

$$\lim_{u \to 0}\left[\left(\frac{i2u}{w-w'}\right)\left(-\frac{2}{ip}e^{(-iw+1/2)ip/2}\frac{(2)^{-iw-1/2}\Gamma(-iw+1/2)}{(mu)^{-iw+1/2}}\right)\left(\frac{2}{ip}e^{-(iw'+1/2)ip/2}\frac{(2)^{iw'-1/2}\Gamma(iw'+1/2)}{(mu)^{iw'+1/2}}\right)\right]$$

$$= \lim_{u \to 0}\left[\left(\frac{i4}{w-w'}\right)\frac{e^{p(w+w')/2}}{mp^2}\left(\frac{2}{m}\right)^{-i(w-w')}\left(\frac{1}{u}\right)^{-i(w-w')}\Gamma(-iw+1/2)\Gamma(iw'+1/2)\right] \tag{54}$$

We can write this result in a more convenient form by noticing that

$$\left(\frac{1}{u}\right)^{-i(w-w')} = \exp\left[\ln\left(\frac{1}{u}\right)^{-i(w-w')}\right] = \exp\left[-i(w-w')\ln\left(\frac{1}{u}\right)\right] \tag{55}$$

Using Euler's identity and rearranging terms, (54) becomes

$$\lim_{u \to 0}\left\{\frac{4i}{mp^2}e^{p(w+w')/2}\left(\frac{2}{m}\right)^{-i(w-w')}\left[\frac{\cos[(w-w')\ln(1/u)]}{w-w'}\right.\right.$$

$$\left.\left. + i\frac{\sin[(w-w')\ln(1/u)]}{w-w'}\right]\Gamma(-iw+1/2)\Gamma(iw'+1/2)\right\} \tag{56}$$

In Appendix A we will show that

$$\lim_{u \to 0} \frac{\cos(w-w')\ln\left(\frac{1}{u}\right)}{w-w'} = 0$$

$$\lim_{u \to 0} \frac{\sin(w-w')\ln\left(\frac{1}{u}\right)}{w-w'} = p\,d(w-w') \tag{57}$$

Using these limits (56) simplifies to

$$\lim_{u \to 0}\left(\frac{i2u}{w-w'} H^{(1)}_{-iw+1/2}(-imu) H^{(1)}_{iw'+1/2}(imu)\right) = -\frac{4}{mp} e^{pw}\Gamma(-iw+1/2)\Gamma(iw+1/2)d(w-w') \tag{58}$$

To obtain the final result, note that the Gamma functions satisfy

$$\Gamma(-iw+1/2)\Gamma(iw+1/2) = |\Gamma(iw+1/2)|^2 = \frac{p}{\cosh pw} \tag{59}$$

This means that we can rewrite (58) in the form

$$\lim_{u \to 0}\left(\frac{i2u}{w-w'} H^{(1)}_{-iw+1/2}(-imu) H^{(1)}_{iw'+1/2}(imu)\right) = -\frac{4e^{pw}}{m\cosh pw} d(w-w') \tag{60}$$

The minus sign in (50) cancels the one here. Moreover, we obtain the exact same result modulo the minus sign for the other term in (48). Adding these results together we find that the normalization is given by

$$\langle y_w | y_{w'} \rangle = \int_0^\infty y_w^\dagger y_{w'}\, du = \frac{8e^{pw}}{m\cosh pw} d(w-w') \tag{61}$$

## 3 Expressing Minkowski wave functions in terms of Rindler wavefunctions

The Minkowski wave function is defined over the entire $(t, z)$ plane. Therefore in order to express it in terms of Rindler wave functions, we need to know the form of the Rindler states in all four regions. We now describe the states in regions II, F, and P and then piece them together to write down the Minkowski wave function.

In order to distinguish the states of regions I, II, F, and P we will adopt the notation $y_w^I, y_w^{II}, y_w^F$, and $y_w^P$ respectively. Furthermore, it is necessary to distinguish between positive and negative frequency modes in the future and past regions. These can be denoted by $y_w^{F(+)}, y_w^{F(-)}, y_w^{P(+)}$, and $y_w^{P(-)}$. Finally, we label the Minkowski wave functions by $y_w^{\pm}$ where $(\pm)$ denotes particle/anti-particle mode respectively.

The derivation and normalization of the Rindler states in the other regions is similar to that used in the previous section. So we simply state the results, continuing to examine the spin-up case. In the future region, the Dirac equation assumes a different form because $u$ becomes a timelike coordinate and $v$ becomes a spacelike coordinate. The spin-up state in the future region is given by[3]

$$y_w^{F(+)} = e^{-iwv} \begin{pmatrix} \Phi_w^- - i\Phi_w^+ \\ 0 \\ \Phi_w^- + i\Phi_w^+ \\ 0 \end{pmatrix} \quad (62)$$

In this case the components of the wave function are given by Hankel functions of the second kind

$$\Phi_w^{\pm} = H_{iw\pm1/2}^{(2)}(mu) \quad (63)$$

The negative frequency state is found to be

$$y_w^{F(-)} = e^{-iwv} \begin{pmatrix} \tilde{f}_w^- - i\tilde{f}_w^+ \\ 0 \\ \tilde{f}_w^- + i\tilde{f}_w^+ \\ 0 \end{pmatrix} \quad (64)$$

where $\tilde{f}_w^{\pm} = H_{iw\pm1/2}^{(1)}(mu)$. Using the procedure outlined in section 2, the student can verify that the normalization of the states in the future region is given by

$$\langle y_w^{F(l)} | y_w^{F(k)} \rangle = \frac{16 e^{-k\pi w}}{m} d(w-w') d_{kl} \quad (k,l = \pm 1) \quad (65)$$

The states in regions P and II are related to the states in regions F and I in the following way

$$y_w^{P(\pm)}(t,z) = a_3 y_w^{F(\mp)}(-t,-z), \quad y_w^{II}(t,z) = a_3 y_w^{I}(-t,-z) \quad (66)$$

As an aside, note that strictly speaking it is necessary to write down the wave functions so that they are only defined in the appropriate region. This can be done using the Heaviside step function. For our purposes, it won't be necessary to worry about this explicitly. We simply note that each Rindler wave function is defined in its given region and is zero elsewhere.

Now let's turn to the problem of expressing the Minkowski states in terms of the Rindler states. In a previous work [3] a complex argument involving source terms on the light cone (which is bound to confuse the student) is used to suggest the form of the Minkowski wave function. We instead take a more heuristic approach based on simple mathematical arguments.

Observing that the Minkowski modes must cover the entire $(t, z)$ plane, we construct them by patching together the Rindler wave functions $y_w^I, y_w^{F(\pm)}, y_w^{P(\pm)},$ and $y_w^{II}$. Then we demand continuity at the boundaries of each region.

To patch together the Rindler wave functions, let's just sum them up with arbitrary constants. A positive frequency Minkowski mode can then be written in terms of the Rindler modes in the following way

$$y_w^+ = N\left(a_1 y_w^I + a_2 y_w^{F(+)} + a_3 y_w^{P(+)} + a_4 y_w^{II}\right) \tag{67}$$

where $N$ is a normalization constant. A negative frequency Minkowski mode is defined similarly

$$y_w^- = M\left(b_1 y_w^I + b_2 y_w^{F(-)} + b_3 y_w^{P(-)} + b_4 y_w^{II}\right) \tag{68}$$

In our simple approach, to determine the values of the $a_i$'s we can compare the wave functions in each region along the boundaries where continuity in Minkowski space requires they match up. For example, consider the Rindler wave functions in regions I and II, which must match up at $u = 0$. We can find the constants by taking the limit in each region as $u \to 0$. This can be demonstrated explicitly by considering the upper component of each spinor.

In the following we use $y_w^{I(U)}$ and $y_w^{II(U)}$ to denote the upper component of the spinors in regions I and II respectively. In the case of region I, the reader can refer to (29) to recall that the upper component of the spinor is given by

$$y_w^{I(U)} = H_{iw-1/2}^{(1)}(imu) + H_{iw+1/2}^{(1)}(imu) \quad \text{(upper component)} \tag{69}$$

Using (52) we can write this in terms of the modified Bessel functions of the second kind, giving

$$y_w^{I(U)} = \frac{2}{pi}\left[e^{-(iw-1/2)ip/2} K_{iw-1/2}(mu) + e^{-(iw+1/2)ip/2} K_{iw+1/2}(mu)\right] \tag{70}$$

Recalling the behavior of the Bessel functions for small argument, for small $u$ this becomes

$$y_w^{I(U)} \approx \frac{2}{pi} e^{pw/2}\left[e^{ip/4} \frac{2^{iw-3/2}\Gamma(iw-1/2)}{(mu)^{iw-1/2}} + e^{-ip/4} \frac{2^{iw-1/2}\Gamma(iw+1/2)}{(mu)^{iw+1/2}}\right] \tag{71}$$

To determine the form of the wave function in region II for small $u$, we apply (66) together with the original definition of the coordinates as given in (1). The $u,v$ coordinates in this region are given by

$$v = \text{arctanh}(z/t), \quad u = \text{sgn}(z)\sqrt{z^2 - t^2} \tag{72}$$

From this we see that setting $z \to -z, t \to -t$ leaves $v$ unchanged while $u \to -u$. We find that the upper component of the spinor in region II as $u \to 0$ is given by

$$y_w^{II(U)} \approx \frac{2}{pi} e^{pw/2} e^{ip/4} \left[ \frac{2^{iw-3/2}\Gamma(iw-1/2)}{(-mu)^{iw-1/2}} + i\frac{2^{iw-1/2}\Gamma(iw+1/2)}{(-mu)^{iw+1/2}} \right] \tag{73}$$

Now, for the (+) frequency mode, we have $-1 = e^{ip}$. Therefore we can proceed as follows

$$y_w^{II(U)} \approx \frac{2}{pi} e^{pw/2} e^{ip/4} \left[ (-1)^{-iw+1/2} \frac{2^{iw-3/2}\Gamma(iw-1/2)}{(mu)^{iw-1/2}} + i(-1)^{-iw-1/2} \frac{2^{iw-1/2}\Gamma(iw+1/2)}{(mu)^{iw+1/2}} \right]$$

$$= \frac{2}{pi} e^{3pw/2} e^{ip/4} \left[ i\frac{2^{iw-3/2}\Gamma(iw-1/2)}{(mu)^{iw-1/2}} + \frac{2^{iw-1/2}\Gamma(iw+1/2)}{(mu)^{iw+1/2}} \right] \tag{74}$$

Comparison with the expression we obtained for the wave function in region I (71) as $u \to 0$ shows that the following relationship must hold

$$-ie^{-pw} y_w^{II} = y_w^{I} \tag{75}$$

A similar procedure can be used to compare the states in the future and past regions to that in region I. We find that

$$y_w^{F(+)} = -e^{-pw/2} e^{-ip/4} y_w^{I}, \quad y_w^{P(+)} = e^{pw/2} e^{ip/4} y_w^{I} \tag{76}$$

Looking back at the expression used for the Minkowski wave function (67) and taking $a_1$ to be unity, (75) and (76) tell us that the positive frequency Minkowski wave function should be written as

$$y_w^{+} = N\left(y_w^{I} - e^{pw/2} e^{ip/4} y_w^{F(+)} + e^{-pw/2} e^{-ip/4} y_w^{P(+)} - ie^{-pw} y_w^{II}\right) \tag{77}$$

The normalization constant can be determined by requiring that

$$\langle y_w^{+} | y_{w'}^{+} \rangle = d(w - w') \tag{78}$$

Using (61) and (65) together with (66) and (77) and (78), we find that

$$N = \sqrt{m/48} \tag{79}$$

A similar procedure can be used to show that the normalization constant for the negative frequency modes is the same and that the Minkowski can be written as

$$y_w^- = \sqrt{m/48}\left(e^{-pw}y_w^I + e^{-pw/2}e^{ip/4}y_w^{F(-)} + e^{-pw/2}e^{-ip/4}y_w^{P(-)} + iy_w^{II}\right) \tag{80}$$

## 4 Calculating the Bogolubov coefficients

In this section we will determine the Bogolubov coefficients which allow us to write the Rindler wave function in terms of positive and negative frequency Minkowski modes. These were originally stated by Soffel[3], and can be obtained by simple algebra. The key concept in this derivation is that the Rindler observer in region I is causally disconnected from region II and vice versa. If we focus on an observer in region I, this means that we seek a combination of $y_w^+$ and $y_w^-$ that reflects this fact by eliminating $y_w^{II}$.

Let's denote the wave function as seen by this Rindler observer in region I by ${}^R y_w^I$. The linear combination of $y_w^+$ and $y_w^-$ we seek is

$$y_w^{R(I)} = e^{pw}y_w^+ + y_w^- = e^{pw/2}\left(e^{pw/2}y_w^+ + e^{pw/2}y_w^-\right) \tag{81}$$

It is necessary to normalize this state. Using (78) we find that

$$\begin{aligned}\langle {}^R y_w^I | {}^R y_{w'}^I \rangle &= e^{p(w+w')/2}\left[e^{p(w+w')/2}\langle y_w^+ | y_{w'}^+\rangle + e^{-p(w+w')/2}\langle y_w^- | y_{w'}^-\rangle\right] \\ &= 2e^{pw}\cosh(pw)d(w-w')\end{aligned} \tag{82}$$

Adding one final layer of notation, we denote the properly normalized Rindler wave function by $\tilde{y}_w^{R(I)}$. Using (82) this is written as

$$\begin{aligned}\tilde{y}_w^{R(I)} &= \frac{1}{\sqrt{2e^{pw}\cosh(pw)}}\,{}^R y_w^I \\ &= \frac{1}{\sqrt{2e^{pw}\cosh(pw)}}\left[e^{pw/2}\left(e^{pw/2}y_w^+ + e^{pw/2}y_w^-\right)\right] \\ &= \frac{e^{pw/2}}{\sqrt{2\cosh(pw)}}y_w^+ + \frac{e^{-pw/2}}{\sqrt{2\cosh(pw)}}y_w^- \\ &= a_w y_w^+ + b_w y_w^-\end{aligned} \tag{83}$$

The Bogolubov coefficients are given by

$$a_w = \frac{e^{pw/2}}{\sqrt{2\cosh(pw)}}, \quad b_w = \frac{e^{-pw/2}}{\sqrt{2\cosh(pw)}} \tag{84}$$

A similar procedure can be applied to the state in the left wedge, by noting that the observer in region II is causally disconnected from region I. This is a good exercise for the student.

## 5 Conclusion

This paper has provided a pedagogical introduction to working with massive Dirac particles in Rindler space. In addition, we have applied the Lagrange-Green identity which greatly simplifies the problem of normalizing the states, obtaining the exact result found in the literature. To the knowledge of the author this approach has not be used before. We concluded by applying simple mathematical arguments to express the Minkowski wave functions in terms of the Rindler states.

The "machinery" described in this paper can be applied to further research. In particular, it can be used to give a full relativistic treatment to problems in quantum information theory that involve massive Dirac particles and an accelerated observer. We intend to pursue this line of research in a future paper. The value of doing so has been demonstrated by recent interest expressed in the properties of entangled particles in a gravitational field [9].

## Appendix A: Mathematical Preliminaries

We now prove three crucial propositions that were used to calculate the normalization of the Rindler states. In the following, we will consider functions $f(x)$ on the real numbers that are infinitely differentiable and that vanish outside of some region. These functions are called *test functions* are denoted by writing $f(x) \in C_0^\infty(\mathbb{R})$.

We begin by considering the well-known *Riemann-Lebesgue lemma*, which basically states that the Fourier coefficients $\hat{f}_k$ of a periodic integrable function vanish as $k \to \infty$.

**Proposition**

$\lim_{k \to \infty} \sin kx = 0$ in the distribution sense.

**Proof**

Let $f(x) \in C_0^\infty(\mathbb{R})$ with support contained in $[-R, R]$. Then

$$\int_{-\infty}^{\infty} \sin kx \, f(x) \, dx = \left(\frac{-\cos kx}{k}\right) f(x) \Big|_{-R}^{R} - \int_{-R}^{R} \left(\frac{-\cos kx}{k}\right) f'(x) \, dx$$

$$= \frac{1}{k} \int_{-R}^{R} \cos kx \, f'(x) \, dx$$

Now, since $\left| \int_{-R}^{R} \cos kx \, f'(x) \, dx \right| \leq \int_{-R}^{R} |\cos kx| |f'(x)| \, dx \leq \int_{-R}^{R} |f'(x)| \, dx$ and $\frac{1}{k} \to 0$ as $k \to \infty$,

we conclude that $\int_{-\infty}^{\infty} \sin kx \, f(x) \, dx \to 0$ as $k \to \infty$, that is, $\lim_{k \to \infty} \sin kx = 0$ in the distribution sense. □

With this result in hand, we can show that $\lim_{k \to \infty} \frac{\sin kx}{x} = p \, d(x)$, which we do in the next proof.

**Proposition**

$\lim_{k \to \infty} \frac{\sin kx}{x} = p \, d(x)$ in the distribution sense.

**Proof**

Let $f(x) \in C_0^{\infty}(\mathbb{R})$ with support contained in $[-R, R]$. Using a Taylor expansion it follows that

$$f(x) \approx f(0) + x y(x)$$

where $y(x)$ is a continuous, infinitely differentiable function that may not have compact support. Using this expansion we have

$$\int_{-\infty}^{\infty} \frac{\sin kx}{x} f(x) \, dx = \int_{-R}^{R} \frac{\sin kx}{x} f(x) \, dx = \int_{-R}^{R} \frac{\sin kx}{x} [f(0) + x y(x)] \, dx$$

$$= f(0) \int_{-R}^{R} \frac{\sin kx}{x} \, dx + \int_{-R}^{R} \sin kx \, y(x) \, dx$$

From the Riemann-Lebesgue lemma, we know that the second term vanishes as $k \to \infty$. Therefore, we only need consider the first term, but it is well known that

$$\int_{-R}^{R} \frac{\sin kx}{x} \, dx = \int_{-kR}^{kR} \frac{\sin y}{y} \, dy \to p \text{ as } k \to \infty. \text{ Therefore we find that}$$

$$\lim_{k \to \infty} \int_{-\infty}^{\infty} \frac{\sin kx}{x} f(x) \, dx = p \, f(0) = \int_{-\infty}^{\infty} p \, d(x) f(x) \, dx$$

That is, $\lim_{k \to \infty} \dfrac{\sin kx}{x} = p\,d(x)$ in the distribution sense. □

The final result we need is to show that $\dfrac{\cos kx}{x}$ vanishes as $k \to \infty$. The proof is similar to the one we just did with some minor modifications.

**Proposition**

$$\lim_{k \to \infty} P.V. \dfrac{\cos kx}{x} = 0$$

in the distribution sense, where P.V. is "principal value".

**Proof**

Let $f(x) \in C_0^\infty(\mathbb{R})$ with support contained in $[-R, R]$. Following the last example, we expand in Taylor $f(x) \approx f(0) + x y(x)$ where $y(x)$ is a continuous, infinitely differentiable function that may not have compact support. Now, by the definition of $P.V. \dfrac{\cos kx}{x}$ we have

$$P.V. \int_{-\infty}^{\infty} \dfrac{\cos kx}{x} f(x)\, dx = P.V. \int_{-R}^{R} \dfrac{\cos kx}{x} f(x)\, dx$$

$$= \lim_{e \to 0^+} \int_{e < |x| < R} \dfrac{\cos kx}{x} f(x)\, dx = \lim_{e \to 0^+} \int_{e < |x| < R} \dfrac{\cos kx}{x} \left[ f(0) + x y(x) \right] dx$$

$$= \lim_{e \to 0^+} f(0) \int_{e < |x| < R} \dfrac{\cos kx}{x}\, dx + \lim_{e \to 0^+} \int_{e < |x| < R} (\cos kx) y(x)\, dx$$

Since $\dfrac{\cos kx}{x}$ is an odd function, $\int_{e < |x| < R} \dfrac{\cos kx}{x}\, dx = 0$. On the other hand, the integrand $(\cos kx) y(x)$ is regular at $x = 0$, hence

$$\lim_{e \to 0^+} \int_{e < |x| < R} (\cos kx) y(x)\, dx = \int_{-R}^{R} (\cos kx) y(x)\, dx$$

However, if we let $k \to \infty$ and use the Riemann-Lebesgue lemma, we conclude that

$$\lim_{k \to \infty} P.V. \int_{-\infty}^{\infty} \dfrac{\cos kx}{x} f(x)\, dx = \lim_{k \to \infty} \int_{-R}^{R} (\cos kx) y(x)\, dx = 0$$

That is, $\lim_{k \to \infty} P.V. \dfrac{\cos kx}{x} = 0$ in the distribution sense. □

The results in derived in this appendix lead to the two results used directly in the paper, that is

$$\lim_{u \to 0} \frac{\cos(w-w') \ln \frac{1}{u}}{w-w'} = 0$$

$$\lim_{u \to 0} \frac{\sin(w-w') \ln \frac{1}{u}}{w-w'} = p\, d(w-w')$$

## Appendix B: The Lagrange-Green Identity

The discussion in this section is from Stackgold (1998) [10]. From elementary differential equations we recall that a *Green's formula* can be used to express an integral in terms of its boundary conditions by using integration by parts. Our goal is to exploit this procedure so that we can evaluate an inner product by inspection.

Consider a second order differential operator $L$ with coefficients $a_i(x)$ which are continuous twice differentiable functions of a real variable that we are denoting by $x$. For the most general case, we can write such an operator as

$$L = a_2(x) \frac{d^2}{dx^2} + a_1(x) \frac{d}{dx} + a_o(x) \tag{85}$$

Next, consider the following integral

$$\int_a^b f(x) L g(x)\, dx = \int_a^b \left( f(x) a_2(x) \frac{d^2 g}{dx^2} + f(x) a_1(x) \frac{dg}{dx} + f(x) a_o(x) g(x) \right) dx \tag{86}$$

where $f(x)$ and $g(x)$ are continuous, twice differentiable functions of a real variable $x$. Integration by parts can transform this expression into

$$\int_a^b f(x) L g(x)\, dx = J(f,g)\Big|_a^b + \int_a^b g(x) L^\dagger f(x)\, dx \tag{87}$$

This is Green's formula. $J$ is known as the *conjunct* of the functions $f$ and $g$ and given (85) can be written as

$$J(f,g) = a_2 \left[ g \frac{df}{dx} - f \frac{dg}{dx} \right] + \left[ a_1 - \frac{da_2}{dx} \right] f g \tag{88}$$

$L^\dagger$ is the adjoint of $L$ which in general is found to be

$$L^\dagger = a_2 \frac{d^2}{dx^2} + \left(2\frac{da_2}{dx} - a_1\right)\frac{d}{dx} + \left(\frac{d^2 a_2}{dx^2} - \frac{da_1}{dx} + a_o\right) \qquad (89)$$

Taking the derivative of (87) with respect to $b$ and then setting $b = x$ gives us Lagrange's identity

$$g L f = \frac{dJ}{dx} + f L^\dagger g \qquad (90)$$

Taken together (87) and (90) are sometimes known as the Lagrange-Green identity. When a differential operator shows up in an inner product we can use this result to transform the integral into one involving a total derivative, i.e.

$$\int \left(g L f - f L^\dagger g\right) dx = \int \frac{dJ}{dx} dx = J(f,g) \qquad (91)$$

### References

1. S. W. Hawking, Nature 248, 30 (1974).

2. W. G. Unruh, Phys. Rev. D14, 870 (1976).

3. M. Soffel, B. Muller, and W. Greiner, Phys. Rev. D22, 1935 (1980).

4. P. M. Alsing and G. J. Milburn, Phys.Rev.Lett. 91, 180404 (2003).

5. N. D. Birrell and P. C. W. Davies, Quantum Fields in Curved Space (Cambridge University Press, N. Y., 1982).

6. R. D'Inverno, Introducing Einstein's Relativity (Oxford University Press, 1992).

7. D. McMahon, Relativity Demystified (McGraw-Hill, 2005).

8. N.N. Lebedev, Special Functions and Their Applications (Dover, 1972).

9. H. Terashima and M. Ueda, Phys. Rev. A 69, 032113 (2004).

10. I. Stakgold, Green's Functions and Boundary Value Problems (Wiley, 1998).